\documentclass[onecolumn,12pt]{IEEEtran}

\IEEEoverridecommandlockouts
\usepackage{textcomp}
\usepackage{stfloats}
\usepackage{verbatim}
\usepackage{cite}
\usepackage{amsmath,amssymb,amsfonts}
\usepackage{algorithmic}
\usepackage{graphicx}
\usepackage{float}
\usepackage{textcomp}
\usepackage{xcolor}
\usepackage{bbm}
\usepackage{array}
\usepackage{rotating}
\usepackage{gensymb}
\usepackage{comment}
\usepackage[acronym]{glossaries}
\usepackage{multirow}
\usepackage{lscape}
\usepackage{makecell}
\usepackage[inline]{enumitem}
\usepackage{svg}
\usepackage{hyperref}
\hypersetup{hidelinks}
\usepackage[font=small,labelsep=period,justification=centering]{caption}
\usepackage{subcaption}
\usepackage{amsmath}

\usepackage{adjustbox}
\usepackage{tikz}
\usetikzlibrary{shapes,arrows}
\usetikzlibrary{positioning}
\usetikzlibrary{decorations.text}

\newcounter{CorrCounter}
\newcounter{LemmaCounter}

\def\BibTeX{{\rm B\kern-.05em{\sc i\kern-.025em b}\kern-.08em
    T\kern-.1667em\lower.7ex\hbox{E}\kern-.125emX}}

\allowdisplaybreaks

\begin{document}

\title{Point Target Near-Field Bistatic Imaging: Chirp-Based Aliasing Analysis}

\author{\IEEEauthorblockN{Baptiste Sambon, Gilles Monnoyer, Claude Oestges, and Luc Vandendorpe \thanks{Baptiste Sambon is a Research Fellow of the Fonds de la Recherche Scientifique - FNRS.}}\\
\IEEEauthorblockA{ICTEAM, UCLouvain - Louvain-la-Neuve, Belgium\\}
email: firstname.lastname@uclouvain.be
}

\maketitle



\begin{abstract}

This paper presents a chirp-based framework for characterising aliasing in a bistatic Near-Field (NF) imaging system equipped with multidimensional antenna arrays. Extending monostatic formulations, we derive closed-form expressions for the maximum spatial frequency, enabling the analytical derivations of the conditions for aliasing-free image reconstruction.
The framework also provides a geometric interpretation of aliasing based on the antenna array geometry, target position, and antenna element spacing. Numerical results corroborate theoretical findings and show that the aliasing-free region enlarges with smaller antenna spacing, greater target range, lower array dimensionality, and smaller arrays. These results enable more effective design of bistatic NF imaging systems.

\end{abstract}



\section{Introduction}
\label{sec:introduction}
 
Integrated Sensing and Communication (ISAC) is emerging as a key enabler for sixth-generation (6G) wireless systems, aiming to unify radar and communication capabilities \cite{chen_6g_2024, liu_integrated_2022}. In this context, the trend towards increasing both the array size and the carrier frequency \cite{8999605} lead to an increase of the Fraunhofer distance and an extension the Near-Field (NF) region, where it is required to account for the Spherical Wavefront (SW) propagation \cite{guerra_near-field_2021, liu_near-field_2023, ye_extremely_2024}. The resulting SW propagation enables accurate spatial localisation beyond conventional angular estimation and also induces a wide spatial spectrum at the receiver\cite{kosasih_spatial_2025}.

To characterise this phenomenon, a chirp-based representation of the steering vector was proposed in \cite{spawc}, enabling the definition of local wave vectors at each antenna. The spatial chirps for the SW regime were shown to well approximate the spatial bandwidth of the received signal. 
This approach provided analytical insights into the spatial sampling requirements at the receiver array for NF point source uplink localisation. It also revealed aliasing effects that introduce additional ambiguities in the localisation ambiguity function, depending on the array geometry and source location. However, the analysis in \cite{spawc} is limited to monostatic configurations with a single point source in an uplink communication scenario and only considers one-dimensional (1D) array geometries. 

In ISAC systems, aliasing introduces interference between spectral components, degrading image quality and target localisation accuracy \cite{zheng2019radar, zhang2022enabling}. 
While $k$-space analysis frameworks exist for mmWave imaging \cite{kazemi_k-space_2021}, they cannot explain key near-field phenomena like aliasing reduction as targets move away from arrays.

Therefore, building on the framework in \cite{spawc}, the present paper extends the chirp-based approach to \textbf{bistatic} \textbf{NF} systems, where the transmit and receive arrays may not be colocated. This extension enables a more complete and rigorous characterisation of \textbf{aliasing} effects in practical ISAC scenarios, in the case of a point scatterer. The analysis is also generalised to handle \textbf{multidimensional arrays}. 

Based on the aforementioned works, the contributions of this paper are summarised as follows.
\begin{itemize}
    \item We derive the conditions for aliasing in bistatic ISAC imaging systems for regular discrete multidimensional arrays.
    \item We leverage the chirp-based approach proposed in \cite{spawc} and generalise it to allow the derivation of closed-form expressions for the spatial sampling requirements to satisfy the non-aliasing conditions.
    \item The chirp-based analysis is applied through numerical analysis to bistatic configurations for a point scatterer, effectively highlighting and analysing the resulting impact on image reconstruction.
    \item Based on the generalised chirp-based approach, a geometric interpretation of the aliasing phenomenon is provided, highlighting how the system parameters influence aliasing patterns in the reconstructed image.
\end{itemize}


\section{System Model}
\label{sec:system_model}


We consider a bistatic configuration where signals emitted by \( N_t \) transmitting antennas are reflected by a point scatterer located at \( \boldsymbol{x}_s \), then captured by \( N_r \) receiving antennas. The transmitter and receiver arrays are respectively defined as  
\begin{equation}
    \mathcal{X}_{\mathrm{A}_t} \triangleq \{ \boldsymbol{x}_t \}_{t=1}^{N_t} \subset \mathbb{R}^d, \quad
    \mathcal{X}_{\mathrm{A}_r} \triangleq \{ \boldsymbol{x}_r \}_{r=1}^{N_r} \subset \mathbb{R}^d,
\label{eq:array_definitions}
\end{equation}
where \( \boldsymbol{x}_t \) and \( \boldsymbol{x}_r \) denote the positions of the transmit and receive antennas. The parameter \( d \in \{1, 2, 3\} \) indicates the spatial dimensionality of the array geometry.
The antenna elements are assumed to be uniformly spaced along each spatial axis, with an inter-element spacing denoted by \( \Delta_{i,m} \), where \( i \in \{x, y, z\} \) indicates the spatial dimension and \( m \in \{r, t\} \) indicates the transmit or receive array.

Following the physical optics approximation \cite{gutierrez-meana_high_2011} and the electromagnetic model of targets in \cite{extended_target}, the noise-free signal at the $r$-th receive antenna from the $t$-th transmit antenna can thus be expressed as 
\begin{equation}
    u(\boldsymbol{x}_{r},\boldsymbol{x}_{t};\boldsymbol{x}_s) = \zeta z(\boldsymbol{x}_s,\boldsymbol{x}_{r}) z(\boldsymbol{x}_s,\boldsymbol{x}_{t}),
\label{eq:signal}
\end{equation}
with 
\begin{equation}
    z(\boldsymbol{x}_s,\boldsymbol{x}_{m}) \triangleq \frac{e^{-jk \| \boldsymbol{x}_s - \boldsymbol{x}_{m} \|}}{\| \boldsymbol{x}_s - \boldsymbol{x}_{m} \|}, \quad m \in \{r, t\}, 
\end{equation}
where \( \zeta \) is a constant that depends on the target surface and \( k~=~\frac{2\pi}{\lambda} \) is the wavenumber with wavelength $\lambda$. In this paper, we assume narrowband signals, which allows us to focus on the wave propagation characteristics of the received signal without considering the temporal aspects. Accordingly, the received signal \( u \) depends solely on the spatial coordinates \( \boldsymbol{x}_{r} \) and \( \boldsymbol{x}_{t} \).


\section{Image Reconstruction Analysis}
\label{sec:ambiguity_function_analysis}


Image reconstruction plays a key role in estimating the point scatterer location \( {\boldsymbol{x}}_s \) from the received signals \( u(\boldsymbol{x}_{r},\boldsymbol{x}_{t};\boldsymbol{x}_s) \), while excluding any artefactual or other responses.  
Reconstruction algorithms typically rely on the result of a matched filtering operation as an initial step to localise or image targets \cite{ oral_3-d_2004, tan_three-dimensional_2017, kidera_tdoa-based_2018, zhang_integrated_2023}. However, aliasing can cause these algorithms to operate in uncontrolled regions where key metrics--resolution, contrast, or accuracy--are no longer guaranteed and may be significantly degraded. 

Assuming initially that the antenna arrays are space-continuous, the reconstructed image $\mathcal{I}$ matched at a \textit{tentative} location \( \tilde{\boldsymbol{x}}_s \) is thus expressed as
\begin{equation}
    \mathcal{I}(\tilde{\boldsymbol{x}}_s ; \boldsymbol{x}_s) = \int_{\mathcal{X}_{\mathrm{A}_{t}}} \int_{\mathcal{X}_{\mathrm{A}_{r}}} u(\boldsymbol{x}_{r},\boldsymbol{x}_{t};\boldsymbol{x}_s)  z^{*}(\tilde{\boldsymbol{x}}_s,\boldsymbol{x}_{r})  z^{*}(\tilde{\boldsymbol{{x}}}_s,\boldsymbol{x}_{t}) \, \mathrm{d}\boldsymbol{x}_r \, \mathrm{d}\boldsymbol{x}_t.
\label{eq:reconstructed_image_continuous}
\end{equation}
For a discrete antenna array, the above integral expressions are replaced by summations over the corresponding antenna elements. However, this spatial discretisation introduces a limitation: if the spacing between adjacent antennas exceeds half the wavelength (i.e., \( \Delta_{i,m} > \lambda/2 \)), spatial aliasing may occur in the reconstructed image \cite{jerri_shannon_1977}. 
This effect can result in repeated artefacts that degrade imaging quality. 



By substituting the expression of \( u(\boldsymbol{x}_{r},\boldsymbol{x}_{t};\boldsymbol{x}_s) \) from \eqref{eq:signal} into the reconstruction formula in \eqref{eq:reconstructed_image_continuous}, the bistatic reconstructed image can be expressed as a product of two separate terms:
\begin{equation}
\mathcal{I}(\tilde{\boldsymbol{x}}_s ; \boldsymbol{x}_s)
= \zeta \, S_t(\tilde{\boldsymbol{x}}_s, \boldsymbol{x}_s) \, S_r(\tilde{\boldsymbol{x}}_s, \boldsymbol{x}_s),
\label{eq:reconstructed_image}
\end{equation}
where 
\begin{equation}
    S_m(\tilde{\boldsymbol{x}}_s, \boldsymbol{x}_s) = \int_{\mathcal{X}_{\mathrm{A}_{m}}} 
    z^{*}(\tilde{\boldsymbol{x}}_s, \boldsymbol{x}_{m}) 
    z(\boldsymbol{x}_s, \boldsymbol{x}_{m}) \, \mathrm{d}\boldsymbol{x}_{m}, \quad m \in \{r, t\}.  
\end{equation}
This expression reveals that the integrals over the transmit and receive arrays are independent. Each term $S_m$ corresponds to a monostatic partial image computed using only one of the two arrays --either the transmitter ($m=t$) or the receiver ($m=r$). The full bistatic image $\mathcal{I}$ is then obtained by multiplying these two partial monostatic contributions. 
This separability also implies that the spatial sampling conditions for the transmit and receive arrays can be analysed independently. Therefore, without loss of generality, we focus in the following on the monostatic terms $S_m$ only and omit the subscript $m$ for clarity.

To derive the aliasing conditions, we must now focus on the chirp structure of the individual terms, i.e. the function
\begin{equation}
    g(\boldsymbol{x} ; \tilde{\boldsymbol{x}}_s, \boldsymbol{x}_s) =  
    z^{*}(\tilde{\boldsymbol{x}}_s, \boldsymbol{x}) \,
    z(\boldsymbol{x}_s, \boldsymbol{x}).
\end{equation}   
For a given spatial wavevector \( \boldsymbol{k} = (k_{x}, k_{y}, k_{z}) \), the Fourier transform $G$ of \( g(\boldsymbol{x} ; \tilde{\boldsymbol{x}}_s, \boldsymbol{x}_s) \) can be expressed as 
\begin{equation}
    G(\boldsymbol{k} ; \tilde{\boldsymbol{x}}_s, \boldsymbol{x}_s) = \int_{\mathcal{X}_\mathrm{A}} g(\boldsymbol{x} ; \tilde{\boldsymbol{x}}_s, \boldsymbol{x}_s) e^{-j \boldsymbol{k}^{T} \cdot \, \boldsymbol{x}} \, \mathrm{d}\boldsymbol{x}_, 
\end{equation}
where \( \boldsymbol{k}^{T} \) is the transpose of \( \boldsymbol{k} \) and $\cdot$ denotes the dot product.

Following the approach in \cite{spawc} for the one-dimensional case, a key observation is that the spectrum \( G \) evaluated at \( \boldsymbol{k}~=~\boldsymbol{0} \) corresponds to the integral of \( g(\boldsymbol{x} ; \tilde{\boldsymbol{x}}_s, \boldsymbol{x}_s) \) over the array aperture and thus to the monostatic reconstructed image $S$, i.e.,
\begin{equation}
    G(\boldsymbol{0} ; \tilde{\boldsymbol{x}}_s, \boldsymbol{x}_s) = \int_{\mathcal{X}_\mathrm{A}} g(\boldsymbol{x} ; \tilde{\boldsymbol{x}}_s, \boldsymbol{x}_s) \, \mathrm{d}\boldsymbol{x} = S(\tilde{\boldsymbol{x}}_s, \boldsymbol{x}_s).
\end{equation}
When discretizing \( \mathcal{X}_\mathrm{A} \) with a spacing of \( \Delta_{i} \) along the \( i \)-th dimension, the spectrum \( G \) becomes periodically replicated along this same dimension with a period of \( \frac{2\pi}{\Delta_{i}} \) \cite{dudgeon_multidimensional_1984}. Such periodic repetitions may introduce aliasing into the reconstructed image if they overlap the region around \( \boldsymbol{k} = \boldsymbol{0} \), where $S$ is encoded. Therefore, the Nyquist sampling criterion \cite{dudgeon_multidimensional_1984, shannon_communication_1949} must be reconsidered in this context: relaxing this criterion by a factor two may still yield accurate reconstruction, as aliasing outside the vicinity of the origin in the spatial frequency domain does not affect the quality of the reconstruction.

The conditions ensuring that a point $\tilde{\boldsymbol{x}}_s$  in the reconstructed image is unaffected by aliasing can thus be expressed as
\begin{equation}
    K_{i}(\tilde{\boldsymbol{x}}_s, \boldsymbol{x}_s) \le 2\pi/\Delta_{i} \quad \forall i \in \{x, y, z\}, 
\label{eq:aliasing_conditions}
\end{equation}
where \( K_{i}(\tilde{\boldsymbol{x}}_s, \boldsymbol{x}_s) \) is the maximum spatial frequency of the spectrum \( G \) along the \( i \)-th dimension. Note that these conditions must be fulfilled for all directions \( i~\in~\{x, y, z\} \) and for \textbf{both} transmit and receive arrays.

Conversely, if the conditions in \eqref{eq:aliasing_conditions} are not satisfied for a given point \( \tilde{\boldsymbol{x}}_s \), aliasing artefacts will appear in the reconstructed image at that location.  A chirp-based interpretation of this maximum spatial frequency is given in the next section.


\section{Chirp-based analysis}
\label{sec:chirp_based_analysis}

The previous section established the conditions under which aliasing appears in the reconstructed image. We now derive a closed-form expression for the maximum spatial frequency along a given axis \( i \), denoted \( K_{i} \), using a chirp-based approach.

A chirp is typically defined as a signal whose frequency varies over time \cite{flandrin_time_2001}. In the present context, the function \( g(\boldsymbol{x} ; \tilde{\boldsymbol{x}}_s, \boldsymbol{x}_s) \) can be interpreted as a spatial chirp, i.e., a signal whose phase varies with the spatial coordinates \( \boldsymbol{x} \) and is parameterised by the target position \( \boldsymbol{x}_s \) and the tentative location \( \tilde{\boldsymbol{x}}_s \). Following the approach in \cite{spawc}, the dominant spatial frequencies in the spectrum are well approximated by  
\begin{equation}
    \left\{ \boldsymbol{k}(\boldsymbol{x} ; \tilde{\boldsymbol{x}}_s, \boldsymbol{x}_s) = \nabla_{\boldsymbol{x}} \phi(\boldsymbol{x} ; \tilde{\boldsymbol{x}}_s, \boldsymbol{x}_s) \right\}, 
\label{eq:stationary_points}
\end{equation}
where \( \phi(\boldsymbol{x} ; \tilde{\boldsymbol{x}}_s, \boldsymbol{x}_s) \) is the phase function of $g$, defined as $\phi(\boldsymbol{x} ; \tilde{\boldsymbol{x}}_s, \boldsymbol{x}_s) \triangleq k \left( \| \boldsymbol{x} - \tilde{\boldsymbol{x}}_s \| - \| \boldsymbol{x} - \boldsymbol{x}_s \| \right)$ with \( k~=~\frac{2\pi}{\lambda} \). 
The expression in \eqref{eq:stationary_points} thus defines the local wavenumber vector \( \boldsymbol{k} \) as the gradient of the chirp phase with respect to the spatial coordinates \( \boldsymbol{x} \), and captures the most present spatial frequency component of the chirp signal at the antenna element location \( \boldsymbol{x} \).


For any position \( \boldsymbol{x} \), its local wavenumber vector \( \boldsymbol{k} \) can be thus be expressed as
\begin{equation}
\begin{aligned}
    \boldsymbol{k}(\boldsymbol{x} ; \tilde{\boldsymbol{x}}_s, \boldsymbol{x}_s)
    &= k \left( \nabla_{\boldsymbol{x}} \| \boldsymbol{x} - \tilde{\boldsymbol{x}}_s \| - \nabla_{\boldsymbol{x}} \| \boldsymbol{x} - \boldsymbol{x}_s \| \right) \\
    &= k \left( \frac{\boldsymbol{x} - \tilde{\boldsymbol{x}}_s}{\| \boldsymbol{x} - \tilde{\boldsymbol{x}}_s \|} - \frac{\boldsymbol{x} - \boldsymbol{x}_s}{\| \boldsymbol{x} - \boldsymbol{x}_s \|} \right). 
\end{aligned}
\label{eq:k_m_}
\end{equation}
    
Each term in \eqref{eq:k_m_} actually corresponds to a local vector of norm $k=\frac{2\pi}{\lambda}$ pointing in the direction between the antenna element located at \( \boldsymbol{x} \) and the points \( \tilde{\boldsymbol{x}}_s \) and \( \boldsymbol{x}_s \), respectively. 


\begin{figure*}[ht]
    \centering
    \begin{tabular}{cccc}
        Monostatic partial image $S_t$ & Monostatic partial image $S_r$ & Bistatic image $\mathcal{I}$ & Zoom  \\
        \begin{subfigure}[t]{0.235\textwidth}
            \centering
            \includegraphics[width=\linewidth, trim=20 20 60 20, clip]{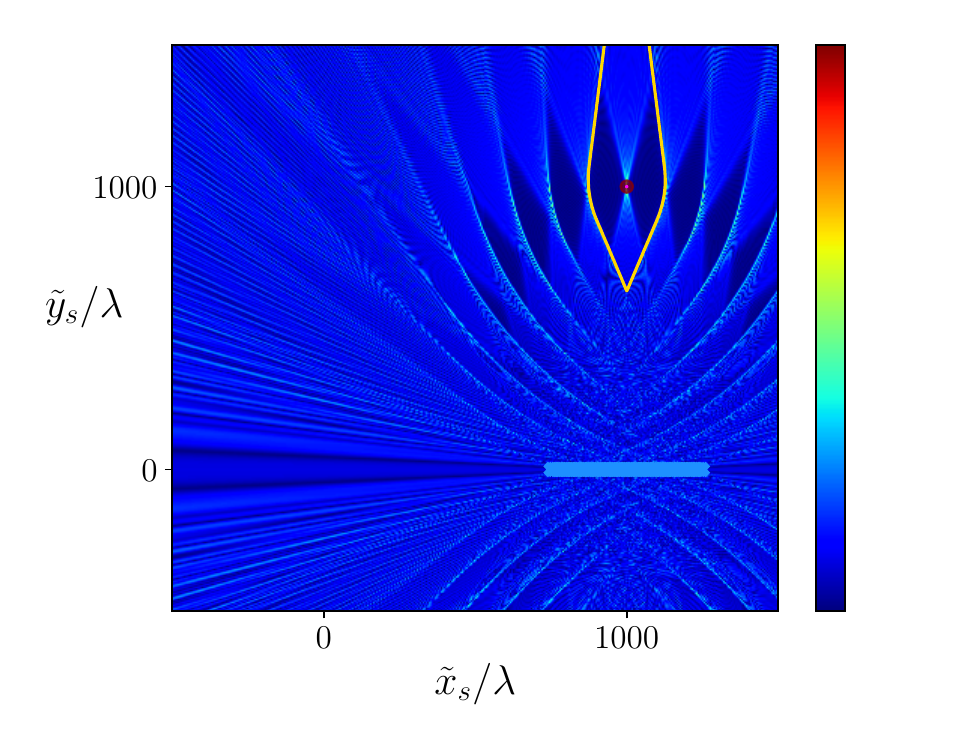}

        \end{subfigure} &
        \begin{subfigure}[t]{0.20\textwidth}
            \centering
            \includegraphics[width=\linewidth, trim=79 20 60 20, clip]{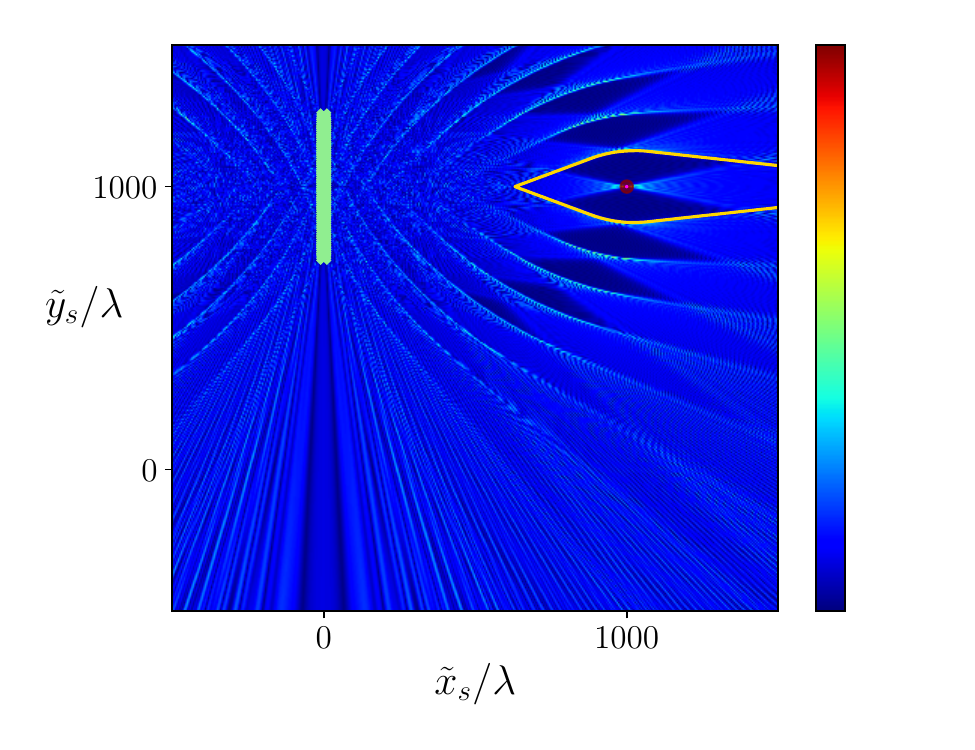}
 
        \end{subfigure} &
        \begin{subfigure}[t]{0.20\textwidth}
            \centering
            \includegraphics[width=\linewidth, trim=79 20 60 20, clip]{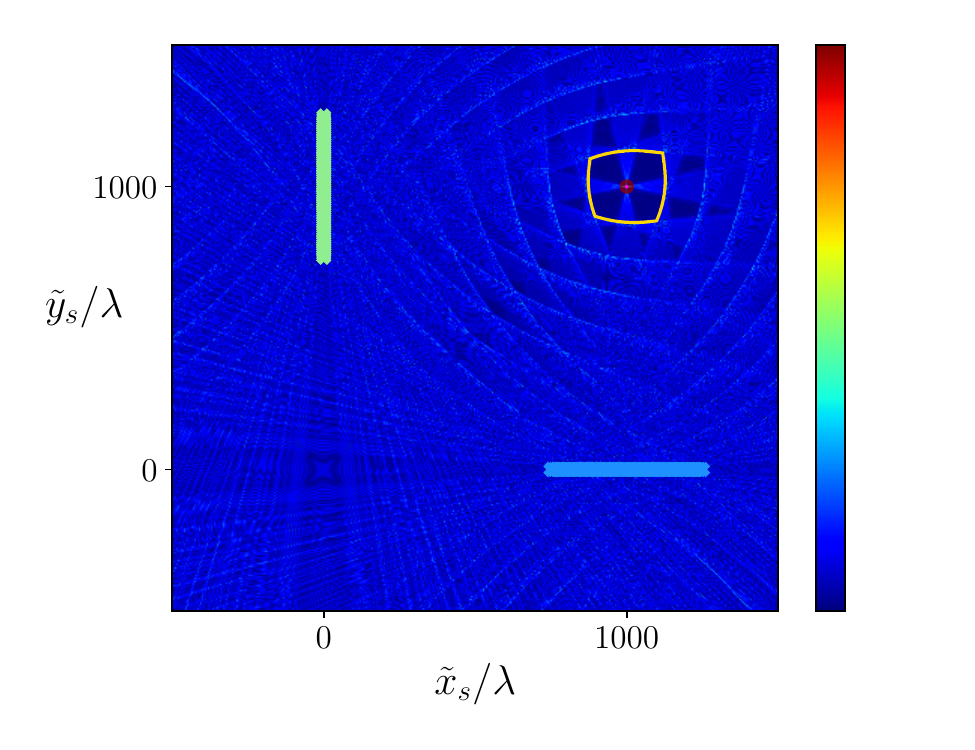}

        \end{subfigure} & 
        \begin{subfigure}[t]{0.22\textwidth}
            \centering
            \includegraphics[width=\linewidth, trim=20 20 80 20, clip]{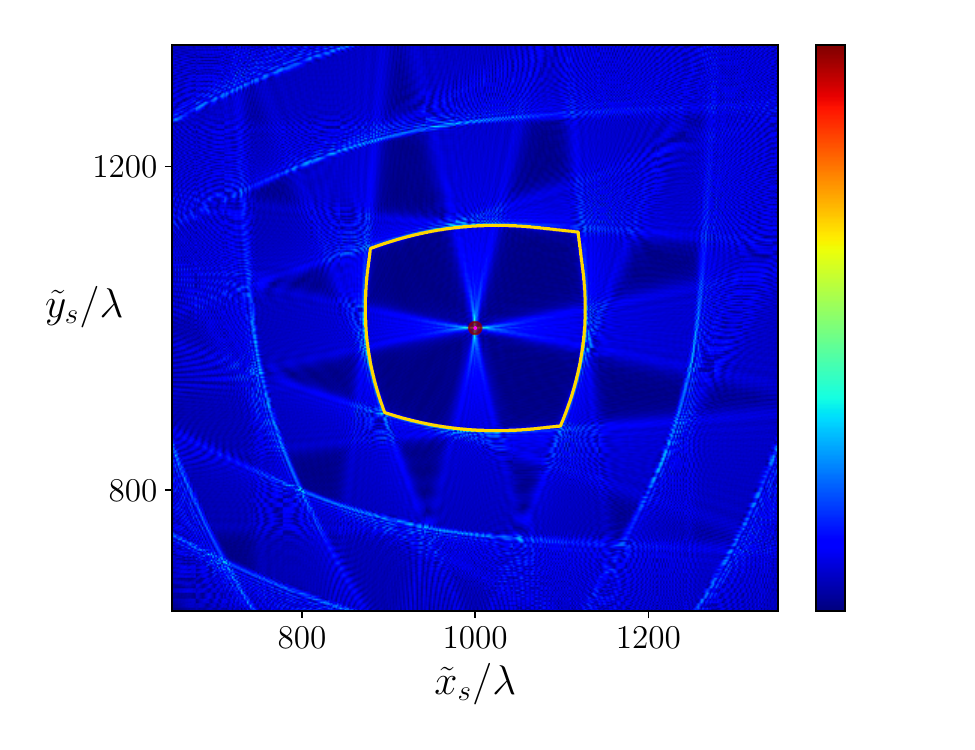}

        \end{subfigure} 
    \end{tabular}

    \includegraphics[width=0.65\linewidth, trim=0 50 0 50, clip]{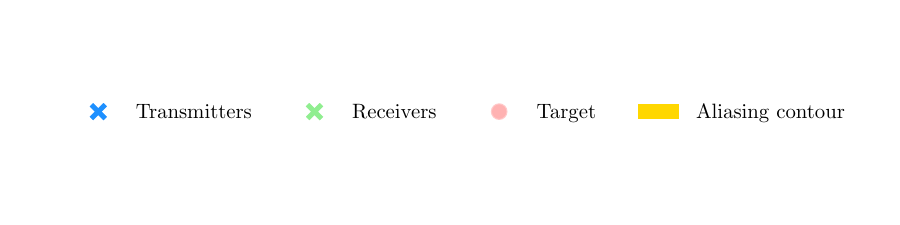}

    \caption{Reconstructed image $\hat{I}(\tilde{\boldsymbol{x}}_s ; \boldsymbol{x}_s)$ for a point scatterer located at $(1000\lambda, 1000\lambda)$ in the $x$-$y$ plane with $500\lambda$ long arrays of $64$ antennas for both transmitter and receiver.. Columns show transmitter term, receiver term, full reconstruction, and aliasing-free region.}
    \label{fig:point_scatterer_aliasing}
\end{figure*}

Using this result, the maximum spatial frequency \( K_{i}(\tilde{\boldsymbol{x}}_s, \boldsymbol{x}_s) \) introduced in \eqref{eq:aliasing_conditions} can be calculated as the maximum value of the $i$-th coordinate of the local wavenumber  $\boldsymbol{k}(\boldsymbol{x} ; \tilde{\boldsymbol{x}}_s, \boldsymbol{x}_s)$, \( k_{i} \), over all array positions \( \boldsymbol{x} \in \mathcal{X}_\mathrm{A} \), i.e.
\begin{equation}
    K_{i}(\tilde{\boldsymbol{x}}_s, \boldsymbol{x}_s) = \max_{\boldsymbol{x} \in \mathcal{X}_\mathrm{A}} \left| k_{i}(\boldsymbol{x} ; \tilde{\boldsymbol{x}}_s, \boldsymbol{x}_s) \right| = \max_{\boldsymbol{x} \in \mathcal{X}_\mathrm{A}} \left| \frac{\partial \phi(\boldsymbol{x} ; \tilde{\boldsymbol{x}}_s, \boldsymbol{x}_s)}{\partial x_{i}} \right|,
\label{eq:K_m_chirp}
\end{equation}
where \( x_{i} \) is the \( i\)-th coordinate of the position vector \( \boldsymbol{x} \).

This maximum spatial frequency \( K_{i} \) along a given dimension \( i \) reflects the strongest spatial variations of the phase function \( \phi(\boldsymbol{x} ; \tilde{\boldsymbol{x}}_s, \boldsymbol{x}_s) \) along this direction among the antenna elements \( \{ \boldsymbol{x} \} \).

To summarise, following the approach proposed in \cite{spawc}, the expression in \eqref{eq:K_m_chirp} provides an effective approximation of the maximum spatial frequency \( K_{i}(\tilde{\boldsymbol{x}}_s, \mathcal{S}) \). Combined with the aliasing conditions in \eqref{eq:aliasing_conditions}, this enables to determine the set of tentative locations \( \tilde{\boldsymbol{x}}_s \) in the reconstructed image that are affected by aliasing, and those that are not.

\section{Numerical Results}
\label{sec:point_scatterer}

This section presents numerical results illustrating how array geometry influences aliasing. A point scatterer is considered, with both the target and arrays (linear or planar) lying in a $x-y$ plane ($d=2$ in \eqref{eq:array_definitions}). Images are reconstructed via matched filtering.

\subsection{Aliasing conditions}

First, we consider a point scatterer located at $(1000\lambda, 1000\lambda)$. The number of antennas is set to \( N_t = N_r = 64 \) over a length of \( 500 \lambda \). The results are shown in Fig.~\ref{fig:point_scatterer_aliasing}.
The first three columns corresponds to transmitter image $S_t$, receiver image $S_r$, and full reconstructed image $\mathcal{I}$, as defined in \eqref{eq:reconstructed_image}. In all columns, a region is outlined in yellow. The interior of this region corresponds to points $\tilde{\boldsymbol{x}}_s$ predicted to be aliasing-free according to \eqref{eq:aliasing_conditions}, using the spatial frequency estimates from \eqref{eq:K_m_chirp}. Outside this outlined region, the presence of aliasing-induced artefacts is expected as the conditions in \eqref{eq:aliasing_conditions} are not satisfied.

\subsubsection{Monostatic partial images}
Due to aliasing, the monostatic partial images associated with the transmit and receive arrays ($S_t$ and $S_r$ in \eqref{eq:reconstructed_image}) alone exhibit repeated peaks along the angular direction in the case of linear antenna arrays, rather than a single one if there was no aliasing. This is consistent with the conventional directional cosine rule observed in FF conditions. In contrast, for NF configurations, the repeating patterns follow a curved conical shape. This highlights the ability of NF imaging to provide accurate spatial localisation beyond mere angular dimension, due to additional inherent range capability of NF systems. 

\subsubsection{Bistatic image}
It can be observed that the reconstructed image $\mathcal{I}$ is indeed obtained as the product of the transmit ($S_t$) and receive ($S_r$) monostatic partial images, according to \eqref{eq:reconstructed_image}. Consequently, a reconstructed point \( \tilde{\boldsymbol{x}}_s \) is affected by aliasing as soon as \textbf{either} the transmit or the receive contribution is aliased. The bistatic aliasing-free region is thus the intersection of the monostatic aliasing-free regions, as shown in Fig.~\ref{fig:point_scatterer_aliasing}.

\subsubsection{Aliasing-free region}
The delimited regions in yellow corresponds closely to the regions where aliasing artefacts are not visible in the reconstructed images. Outside these regions, aliasing manifests as spurious high-magnitude responses that may be mistaken for actual scatterer locations, thereby degrading the reliability of the reconstruction by creating additional ambiguities. 
This aliasing-free region is located around the point scatterer and corresponds to points where \eqref{eq:aliasing_conditions} is respected for all directions and for both transmit and receive antennas. 

\subsection{Parameter Analysis}
\label{sec:parameter_analysis}

\begin{figure}
\centering 

\includegraphics[width=0.4\linewidth, trim=0 8 0 22, clip]{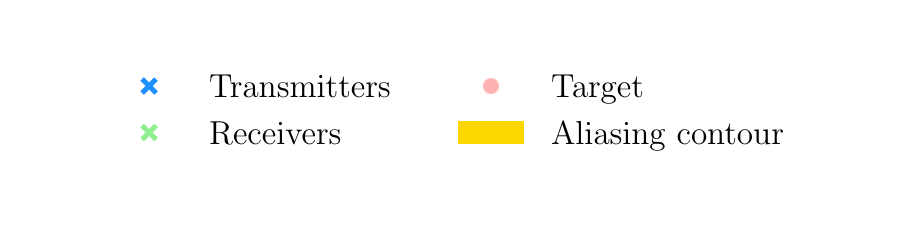}


\begin{subfigure}[t]{0.5\linewidth}
\makebox[0.49\linewidth][c]{$N = 16$}%
\makebox[0.45\linewidth][c]{$N = 64$}
\centering 
\includegraphics[width=0.49\linewidth, trim=20 20 80 15, clip]{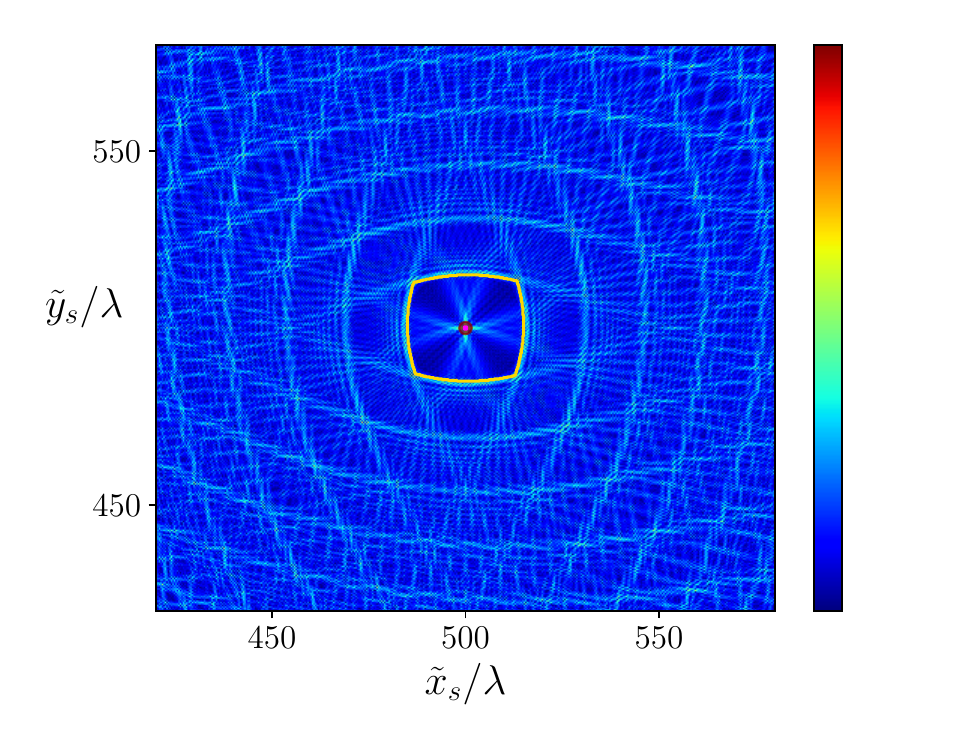}
\includegraphics[width=0.45\linewidth, trim=80 20 50 15, clip]{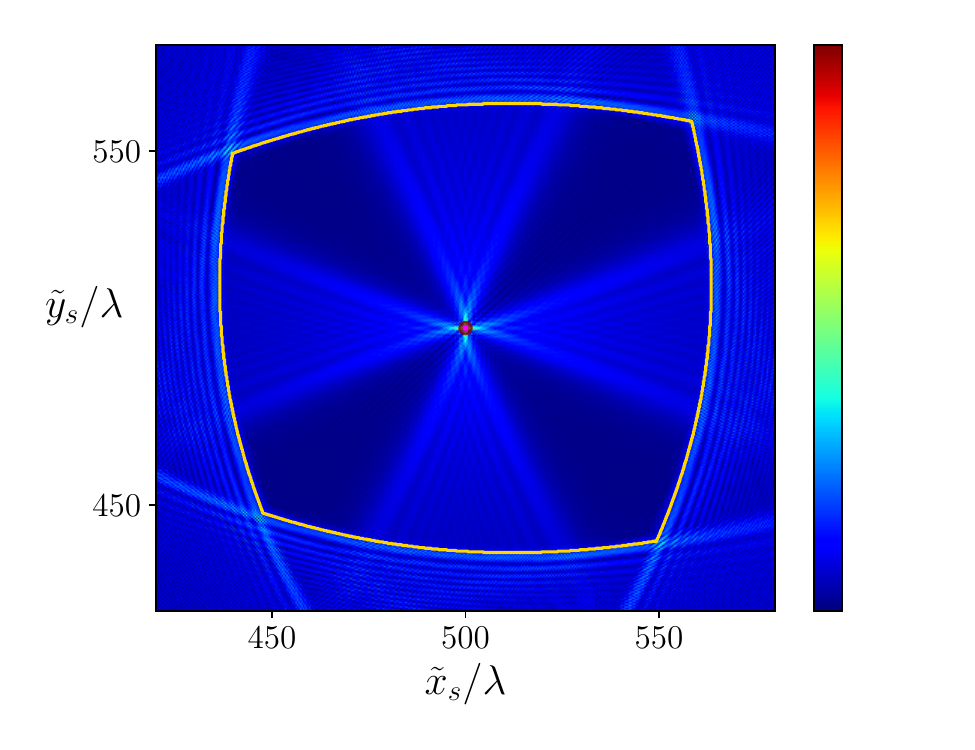}
\caption{Impact of antenna spacing for a fixed array length of $500 \lambda$.} 
\label{fig:impact_spacing}
\end{subfigure}

\vspace{0.2cm}

\begin{subfigure}[t]{0.5\linewidth}
\centering 
\makebox[0.49\linewidth][c]{$N=32, L = 250\lambda$}%
\makebox[0.45\linewidth][c]{$N=128, L = 1000\lambda$}
\includegraphics[width=0.49\linewidth, trim=20 20 80 15, clip]{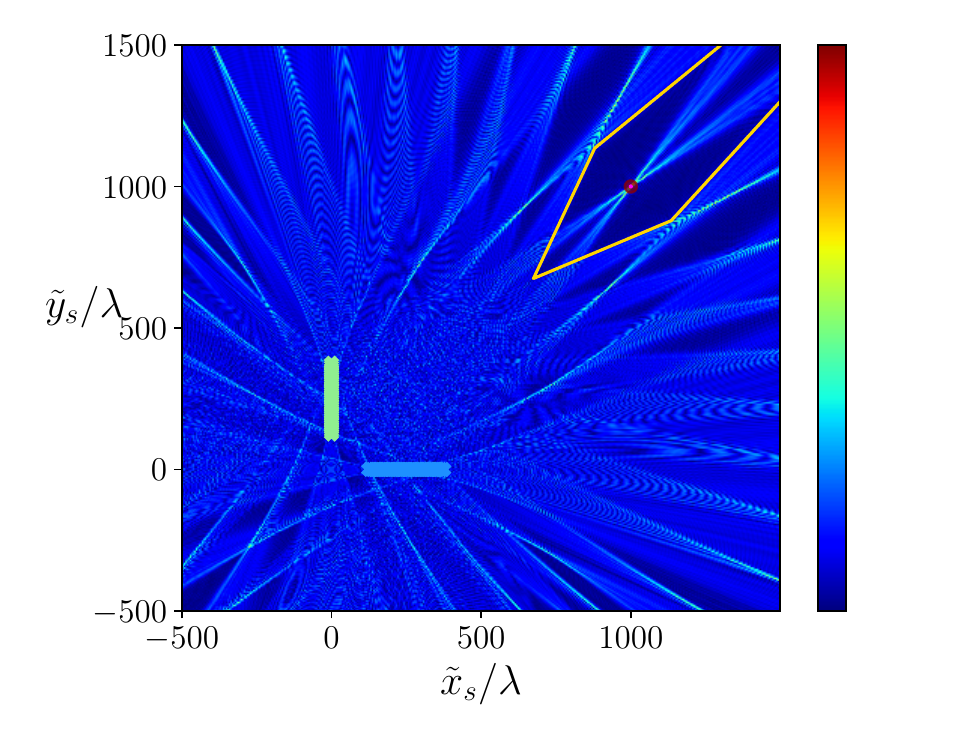}
\includegraphics[width=0.45\linewidth, trim=80 20 50 15, clip]{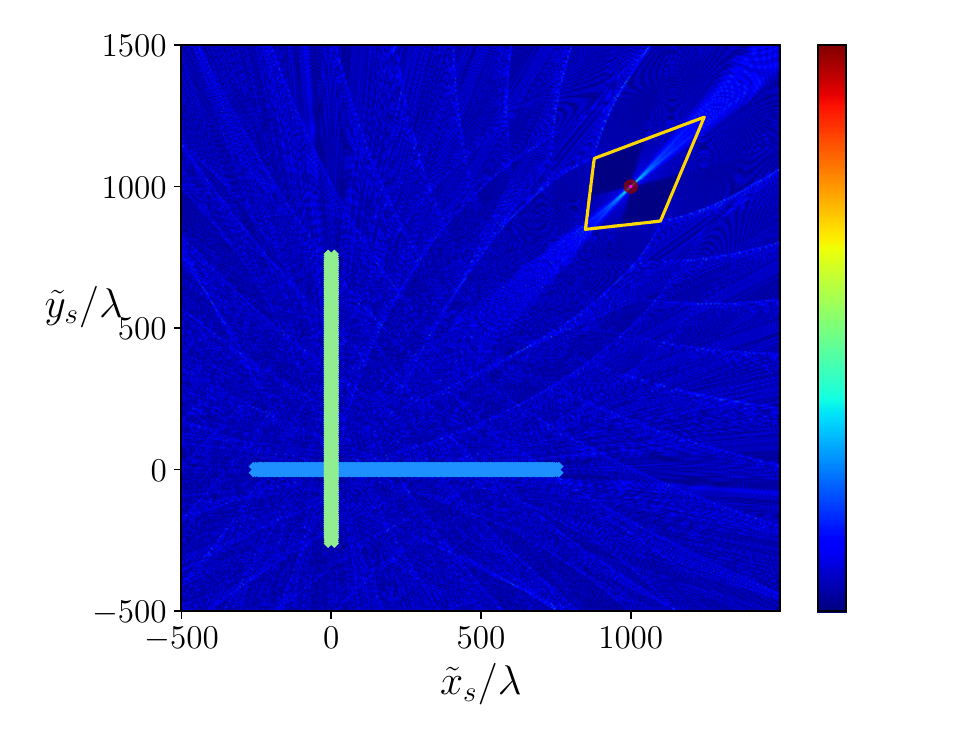}
\caption{Impact of the array length for a fixed antenna spacing.} 
\label{fig:impact_length}
\end{subfigure}

\vspace{0.2cm}

\begin{subfigure}[t]{0.5\linewidth}
\centering
\makebox[0.49\linewidth][c]{$N=64$}%
\makebox[0.45\linewidth][c]{$N=64 \times 64$}
\includegraphics[width=0.49\linewidth, trim=20 20 80 19, clip]{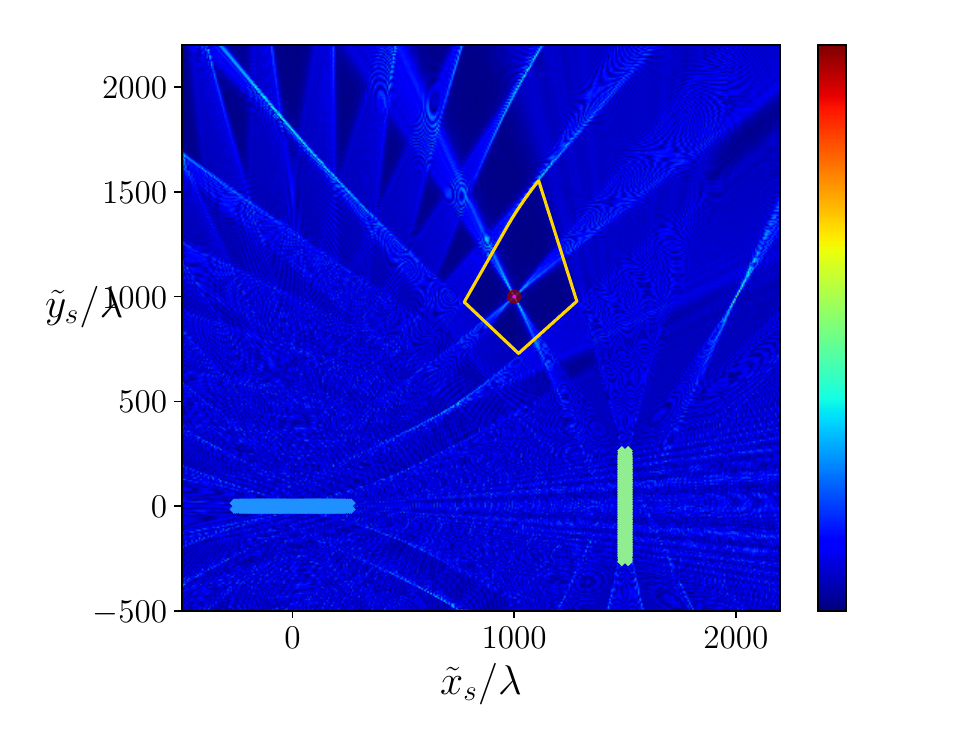}
\includegraphics[width=0.45\linewidth, trim=80 20 50 19, clip]{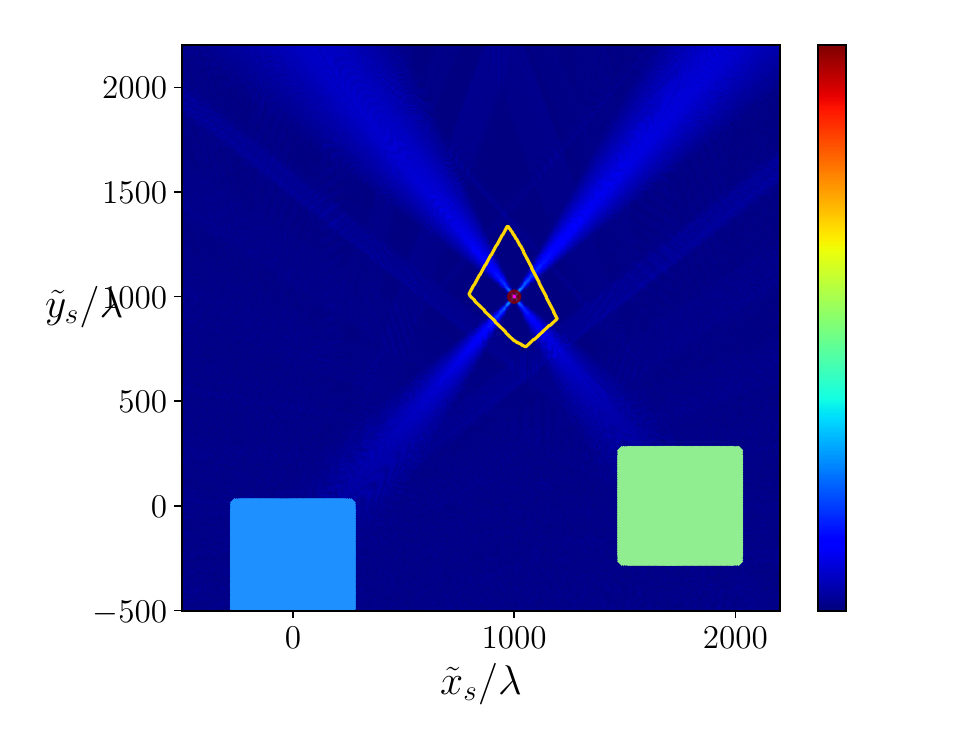}
\caption{Impact of the array dimensionnality.} 
\label{fig:impact_multi_dimensional}
\end{subfigure}

\caption{Impact of system parameters on the reconstructed image for a point scatterer. }

\label{fig:impact_parameters}
\end{figure}

This subsection investigates how system parameters influence aliasing effects, with results shown in Fig.~\ref{fig:impact_parameters}.

\subsubsection{Antenna spacing}
Fig.~\ref{fig:impact_spacing} illustrates antenna spacing impact on reconstructed images. Linear transmit and receive arrays are positioned along the $x$- and $y$-axes, centered at $(500\lambda, 0)$ and $(0, 500\lambda)$ to symmetrically cover the point scatterer at $(500\lambda, 500\lambda)$. The array lengths are fixed at $500\lambda$ with $N = 16$ antennas (left column) and $N = 64$ antennas (right column).
Here, we observe that as the spacing \(\Delta_{i}\) increases (i.e., fewer antennas), the aliasing-free region shrinks. This is because the spatial spectrum \(G\) is repeated more frequently, with a period \(2\pi/\Delta_{i}\) that decreases, increasing the risk of violating the non-aliasing conditions in \eqref{eq:aliasing_conditions}. 

\subsubsection{Array size}
The influence of the array size on aliasing is illustrated in Fig.~\ref{fig:impact_length}. The left column shows the results for a transmit and receive array of length \(250\lambda\), while the right column shows the results for a length of \(1000\lambda\) (same antenna spacing).
Since the antenna spacing is fixed across both plots, the aliasing-free region is not determined by the spectral repetition in \eqref{eq:aliasing_conditions}, but rather by the maximum spatial frequency \( K_i \). As the array size increases, the value of \( K_i \) for a given candidate point $\tilde{\boldsymbol{x}}_s$ remains constant or increases, since the local spatial frequencies $k_i$ in \eqref{eq:K_m_chirp} are evaluated at new antenna positions. This increases the risk of aliasing and reduces the size of the aliasing-free region. 

\subsubsection{Scatterer position}
The maximum spatial frequency component \( K_{i} \) also depends on the scatterer position relative to the arrays. When the scatterer moves closer to the array, the phase variations across the array increase, leading to higher spatial frequencies and a greater risk of aliasing. This is illustrated in the right-most images of Figs.~\ref{fig:point_scatterer_aliasing} and \ref{fig:impact_spacing}, where the aliasing-free region is noticeably larger for a scatterer located at \( 1000\lambda \) compared to one at \( 500\lambda \), under the same antenna array configuration.

\subsubsection{Array dimensionality}
Fig.~\ref{fig:impact_multi_dimensional} illustrates the impact of array dimensionality on aliasing. The left column shows results for linear (1D) transmit and receive arrays, while the right column corresponds to two-dimensional (2D) arrays.
The aliasing-free region is significantly smaller in the 2D case as the aliasing conditions in \eqref{eq:aliasing_conditions} must now be satisfied along multiple spatial dimensions for each array. The number of antenna array elements required to meet the non-aliasing conditions also grows whith the array dimensionality, making the requirements more stringent. 
As a result, the set of spatial locations that satisfy the non-aliasing criteria shrinks, reducing the region in the reconstructed image where aliasing artefacts are absent.
However, extending the arrays to two dimensions enables a more complete sampling of the spatial frequencies of the target, which helps mitigate aliasing artefacts. 

\section{Conclusion}
\label{sec:conclusion}

This paper derived aliasing-free conditions for near-field bistatic imaging systems. We demonstrated that spatial sampling requirements for transmit and receive arrays can be analysed independently and along each dimension. The chirp-based approach provides closed-form expressions for maximum spatial frequencies, enabling aliasing prediction in reconstructed images.
Numerical results corroborated the validity of the derived conditions. Results show aliasing conditions are satisfied in a region around the target where reconstruction is accurate, while degrading artifacts appear outside this region as predicted. Analysis of array geometry and inter-antenna spacing revealed that the aliasing-free region shrinks with increased antenna spacing, larger and/or multidimensional arrays, and closer targets.
Future work will extend the analysis to noisy environments and complex target geometries with non-uniform antenna spacing.


\bibliographystyle{ieeetr}
\bibliography{biblio}

\end{document}